\documentclass[aps,prd,preprintnumbers,groupedaddress,nofootinbib,amssymb,notitlepage,eqsecnum]{revtex4-2}
\usepackage{here}
\usepackage[dvipdfmx]{graphicx}
\usepackage{amsmath,amsthm,amssymb}
\usepackage{bm}
\usepackage{color}

\usepackage{amsfonts}
\usepackage{dcolumn}
\usepackage{hyperref}
\allowdisplaybreaks[1]
\usepackage{stackengine}

\newcommand{\be}{\begin{equation}}  
\newcommand{\ee}{\end{equation}}
\newcommand{\ba}{\begin{eqnarray}}
\newcommand{\ea}{\end{eqnarray}}

\newcommand{\rd}{{\rm d}}

\newcommand{\baf}{\bar{f}}
\newcommand{\bah}{\bar{h}}

\newcommand{\bem}{\begin{bmatrix}}
\newcommand{\eem}{\end{bmatrix}}
\newcommand{\Mpl}{M_{\rm Pl}}

\allowdisplaybreaks

\begin{document}

\preprint{WUCG-23-05}
\title{Black holes in a new gravitational theory 
with trace anomalies}

\author{Shinji Tsujikawa}

\affiliation{
Department of Physics, Waseda University, 3-4-1 Okubo, 
Shinjuku, Tokyo 169-8555, Japan}

\begin{abstract}

In a new gravitational theory with the trace anomaly recently 
proposed by Gabadadze, we study the existence of hairy black hole 
solutions on a static and spherically symmetric background. 
In this theory, the effective 4-dimensional action contains a kinetic 
term of the conformal scalar field related to a new scale $\bar{M}$ 
much below the Planck mass. This property can overcome
a strong coupling problem known to be present in general relativity 
supplemented by the trace anomaly as well as
in 4-dimensional Einstein-Gauss-Bonnet gravity. 
We find a new hairy black hole solution arising from 
the Gauss-Bonnet trace anomaly, which satisfies regular boundary 
conditions of the conformal scalar and metric on the horizon.
Unlike unstable exact black hole solutions with a divergent derivative 
of the scalar on the horizon derived for some related theories 
in the literature, we show that our hairy black hole solution 
can be consistent with all the linear stability conditions of odd- 
and even-parity perturbations.

\end{abstract}

\date{\today}


\maketitle

\section{Introduction}
\label{introsec}

It is well known that the quantum field theory of gravitation gives rise to 
a trace anomaly which is absent at classical level. 
Quantum corrections to the graviton propagator arising from 
loops of massless particles (photons and fermions) were originally 
computed in Refs.~\cite{Capper:1974ed,Capper:1973mv} by using 
a dimensional regularization scheme. 
For the massless field system interacting with gravity, 
Capper and Duff \cite{Capper:1974ic} showed that the conformal 
invariance under Weyl scaling of the metric tensor $g_{\mu \nu}$ 
no longer holds at quantum level. 
In 4-dimensional spacetime, the regularized 
energy-momentum tensor develops nonvanishing trace 
anomalies which consist of curvature scalar quantities 
constructed from Riemann and Ricci tensors \cite{Deser:1976yx}.

The general expression for the gravitational trace anomaly 
derived from one loop calculations in massless theories is given by 
$T_A=-\alpha {\cal G}+\beta W^2+\gamma \square R$, where 
${\cal G}$ is a Gauss-Bonnet (GB) term, $W^2=C_{\mu\nu\rho\sigma}C^{\mu\nu\rho\sigma}$ 
is the squared of a Weyl tensor $C_{\mu\nu\rho\sigma}$, 
and $R$ is a Ricci 
scalar \cite{Duff:1977ay,Birrell:1982ix}.
The constants $\alpha$, $\beta$, $\gamma$ are related to the numbers 
of real scalar fields, Dirac fermions, and vector fields present 
in conform field theory (CFT) \cite{Dowker:1976zf,Christensen:1977jc,Brown:1976wc,Brown:1977pq,Davies:1977ze,Hawking:1976ja}. Indeed, these coefficients exactly coincide  
with those obtained by using the AdS/CFT correspondence in 
strongly coupled large $N$ CFT \cite{Henningson:1998gx}. 
Decomposing  the metric tensor into 
$g_{\mu \nu}=e^{2\sigma}\bar{g}_{\mu \nu}$, 
where $e^{2\sigma}$ is a conformal factor and the metric 
$\bar{g}_{\mu \nu}$ is restricted to have a fixed determinant, 
Riegert \cite{Riegert:1984kt} derived an effective local action 
${\cal S}_A$ that generates the trace anomaly $T_A$ 
in the field equations of motion.
This can be further promoted to consider 
a conformal geometry 
\be
\bar{g}_{\mu \nu}=e^{-2\phi}g_{\mu \nu}\,,
\label{barg}
\ee
without assuming any constraint on  
$\bar{g}_{\mu \nu}$ \cite{Komargodski:2011vj,Fernandes:2021dsb}. 
The Weyl invariant combination 
$\bar{g}_{\mu \nu}=e^{-2\phi}g_{\mu \nu}$
transforms as a metric under diffeomorphisms, with $\phi$ being 
a massless scalar field (dilaton). 
The metric $\bar{g}_{\mu \nu}$ is 
invariant under Weyl transformations 
$g_{\mu \nu} \to e^{2\sigma} g_{\mu \nu}$ and 
$\phi \to \phi+\sigma$.

The diffeomorphism-invariant  effective action ${\cal S}_A$ of gravitational 
trace anomalies contains Galileon-type self-interaction of 
the dilaton and derivative couplings 
to the Einstein tensor besides the couplings between 
$\phi$ and ${\cal G}, W^2$ \cite{Riegert:1984kt,Komargodski:2011vj}. 
The same 4-dimensional action arises from a Wess-Zumino term 
for a gravitational theory nonlinearly realizing the 
conformal symmetry \cite{Gabadadze:2020tvt}. 
Moreover, the action same as the Riegert's one also appears 
in regularized 4-dimensional 
Einstein-Gauss-Bonnet (4DEGB) theory \cite{Glavan:2019inb} after 
a regularized Kaluza-Klein reduction of the $D$-dimensional Einstein-GB gravity with a rescaling of the GB coupling 
constant $\alpha \to \alpha/(D-4)$ \cite{Lu:2020iav,Kobayashi:2020wqy,Fernandes:2022zrq,Tsujikawa:2022lww}.
In 4DEGB gravity, the scalar field $\phi$ corresponds to a radion mode 
characterizing the size of a maximally symmetric internal space. 
We note that the 4-dimensional action generating the GB trace anomaly belongs to a subclass of Horndeski theories with second-order 
field equations of motion \cite{Horndeski,Deffayet:2011gz,Kobayashi:2011nu,Charmousis:2011bf}.

General Relativity (GR) supplemented by the Riegert's action is  
plagued by a strong coupling problem at arbitrary  low energy 
scales \cite{Bonifacio:2020vbk,Gabadadze:2023quw}. 
This is attributed to the fact that the conformal scalar field $\phi$ is not 
a propagating degree of freedom.
In other words, the effective 4-dimensional action does not possess 
a kinetic term with the proper sign.
In 4DEGB gravity, the same strong coupling problem was also 
recognized by studying linear perturbations around static and 
spherically symmetric black holes (BHs) \cite{Tsujikawa:2022lww,Kase:2023kvq} 
and neutron stars (NSs) \cite{Minamitsuji:2022tze}.
In this case the kinetic term associated with the radion 
perturbation $\delta \phi$ vanishes everywhere. 
Moreover, there are ghost/Laplacian instabilities \cite{Tsujikawa:2022lww,Kase:2023kvq} for exact 
BH solutions derived in 4DEGB 
gravity \cite{Lu:2020iav,Fernandes:2021ysi}.

To circumvent the above strong coupling problem, one may add 
a scalar kinetic term to the Riegert's action. 
However, adding terms depending on the scalar field explicitly  
violates the correct structure of gravitational trace anomalies. 
Instead, Gabadadze \cite{Gabadadze:2023quw} recently proposed 
a new gravitational action of the form 
\be
{\cal S}_{R \bar{R}}=M^2 \int \rd^4 x \sqrt{-g}R
-\bar{M}^2 \int \rd^4 x \sqrt{-\bar{g}}\bar{R}\,,
\label{SRbR}
\ee
where $g$ and $\bar{g}$ are the determinants of $g_{\mu \nu}$ and 
$\bar{g}_{\mu \nu}=e^{-2\phi}g_{\mu \nu}$, respectively, and 
$\bar{R}$ is the Ricci scalar evaluated with $\bar{g}_{\mu \nu}$.
The constant $M$ is related to the reduced Planck mass $\Mpl$ as 
$M=\Mpl/\sqrt{2}$, whereas the new mass scale $\bar{M}$ 
is much smaller than $M$. 
The total effective action incorporating the trace anomaly 
is given by ${\cal S}={\cal S}_{R \bar{R}}+{\cal S}_A$, 
where ${\cal S}_A$ is the Riegert's action without 
a fixed determinant constraint on $\bar{g}_{\mu \nu}$.
The Ricci scalar $\bar{R}$ can be expressed in terms of 
the sum of the term $e^{2\phi}R$ and 
derivatives of $\phi$. 
Then, the action (\ref{SRbR}) is equivalent to 
\be
{\cal S}_{R \bar{R}}=\int \rd^4 x \sqrt{-g} 
\left( M^2 R-\varphi^2 R-6 g^{\mu \nu} \nabla_{\mu}\varphi 
\nabla_{\nu} \varphi \right)\,,
\label{SRbR2}
\ee
where $\varphi \equiv \bar{M} e^{-\phi}$ and the covariant 
derivative operator $\nabla_{\mu}$ is associated with the metric 
$g_{\mu \nu}$. Thus, the action ${\cal S}_{R \bar{R}}$ contains the scalar field 
kinetic term $-6 g^{\mu \nu} \nabla_{\mu}\varphi \nabla_{\nu} \varphi$  
with the proper sign\footnote{There might be some alternative mechanisms 
of curing the strong coupling problem, but the fact that the structure of 
trace anomalies is violated by adding explicit $\phi$-dependent terms 
limits those possibilities.}. 
The last two terms in Eq.~(\ref{SRbR2}) respect the conformal 
invariance, while the first Einstein-Hilbert term explicitly breaks
the conformal symmetry.
 
The theory given by the action ${\cal S}={\cal S}_{R \bar{R}}+{\cal S}_A$
can be regarded as an effective field theory (EFT) valid 
below the scale $\bar{M}$. 
Introducing a canonical normalized field $\pi=\bar{M}\phi$, 
we see that nonlinear scalar derivative terms of $\pi$ 
are strongly coupled above the scale $\bar{M}$, 
while they are weakly coupled 
below $\bar{M}~(\ll M)$ \cite{Gabadadze:2023quw}. 
Taking the limit $\bar{M} \to 0$ means that the theory is strongly 
coupled at any scales, as it happens in GR supplemented by the trace anomaly 
and in 4DEGB gravity.

In this paper, we will apply the new gravitational theory of Gabadadze 
to the investigation of static and spherically symmetric BHs 
with scalar hairs. It is anticipated that the presence of 
a canonical scalar kinetic term in ${\cal S}_{R\bar{R}}$ 
as well as the existence of a new scale $\bar{M}$ should allow 
the possibility for overcoming the strong coupling and instability 
problems present for hairy BH solutions in 4DEGB gravity. 
Indeed, we will show that there is a new class of BH solutions 
where the scalar hair arises from the GB trace anomaly. 
If we impose a condition that the two metric components $\bar{f}$ 
and $\bar{h}$ associated with $\bar{g}_{\mu \nu}$
are identical to each other, there is an exact BH solution analogous to those 
derived in 4DEGB gravity \cite{Lu:2020iav,Fernandes:2021ysi} 
and in gravitational theory with a conformal scalar field \cite{Fernandes:2021dsb}. 
However, we will see that the solution consistent with all the field 
equations of motion and regular boundary conditions of 
$\phi$, $\bar{f}$, $\bar{h}$ on the horizon satisfies 
$\bar{f}(\bar{r}) \neq \bar{h}(\bar{r})$ 
at arbitrary distances $\bar{r}$, 
where the difference between 
$\bar{f}$ and $\bar{h}$ comes from the trace anomaly. 
Under the expansion of a small coupling constant $\alpha$, 
we will derive analytic solutions 
to $\bar{f}$, $\bar{h}$, and $\phi$ up to fourth order. 
We will also confirm that they are in very 
good agreement with numerically integrated solutions.

For our new BH solution, the radial field derivative $\phi'(\bar{r})$ is 
a finite constant on the horizon ($\bar{r}=\bar{r}_h$). 
On the other hand, for the exact BH solution present in 4DEGB 
gravity \cite{Lu:2020iav,Fernandes:2021ysi}, 
$\phi'(\bar{r})$ diverges at $\bar{r}=\bar{r}_h$. 
The latter property leads to the linear 
instability of BHs in the vicinity of the horizon \cite{Tsujikawa:2022lww,Kase:2023kvq}. 
This instability was shown for a time-independent 
scalar field in full  Horndeski theories \cite{Minamitsuji:2022vbi} 
(including the shift-symmetric case \cite{Minamitsuji:2022mlv}) 
by using the general results of BH perturbations formulated in Refs.~\cite{Kobayashi:2012kh,Kobayashi:2014wsa,Kase:2021mix}.
We will consider odd- and even-parity perturbations about our new 
BH solution derived under the expansion of the small $\alpha$ 
and show that all the linear stability conditions can be consistently 
satisfied without the strong coupling problem. 
In particular, the propagation speeds of gravitational and 
scalar field perturbations are close to the speed of light with corrections 
induced by the GB trace anomaly. Thus, the new gravitational theory of Gabadadze 
gives rise to a linearly stable BH solution with the scalar hair induced by 
the trace anomaly.

\section{Gravitational action with trace anomalies}
\label{scasec}

In this section, we first briefly review the Riegert's 
action \cite{Riegert:1984kt} and then proceed to the explanation of 
the new gravitational action recently proposed by Gabadadze \cite{Gabadadze:2023quw}.

\subsection{Riegert's action}

As we already mentioned in Introduction, the trace anomaly 
obtained from closed loop calculations for massless fields 
in an external gravitational background has the following 
general expression \cite{Duff:1977ay,Birrell:1982ix} 
\be
T_A=-\alpha {\cal G}+\beta W^2+\gamma \square R\,,
\label{TA}
\ee
where the GB term ${\cal G}$ and the Weyl tensor squared $W^2$ 
are defined by
\ba
{\cal G} &\equiv& R^2-4R_{\mu\nu}R^{\mu\nu}
+R_{\mu\nu\rho\sigma}R^{\mu\nu\rho\sigma}\,,\\
W^2 &\equiv& \frac{1}{3} R^2-2R_{\mu\nu}R^{\mu\nu}
+R_{\mu\nu\rho\sigma}R^{\mu\nu\rho\sigma}\,,
\ea
with $R_{\mu\nu}$ and $R_{\mu\nu\rho\sigma}$ being the 
Ricci tensor and Riemann tensors, respectively.
The coefficients $\alpha$ and $\beta$ are independent of 
the scheme of renormalization, 
while $\gamma$ is not \cite{Dowker:1976zf,Christensen:1977jc,Brown:1976wc,Brown:1977pq,Davies:1977ze,Hawking:1976ja}. 
In this regard, we do not consider the last term of Eq.~(\ref{TA}) and set 
\be
\gamma=0\,,
\ee
in the following discussion.

To derive an action whose variation leads to the trace $T_A$, 
Riegert split the metric tensor $g_{\mu \nu}$ into $g_{\mu \nu}=
e^{2\phi} \bar{g}_{\mu \nu}$ and imposed that the determinant of 
$\bar{g}_{\mu \nu}$ is fixed. 
It is also possible to reformulate the construction of 
the action without putting a constraint on $\bar{g}_{\mu \nu}$. 
In this case, the metric tensor $\bar{g}_{\mu \nu}$ plays 
a dynamical role.
Indeed, the Riegert's action can be reconstructed only by  
requiring the conformal invariance under simultaneous Weyl transformations 
$g_{\mu \nu} \to e^{2\sigma} g_{\mu \nu}$ and 
$\phi \to \phi+\sigma$  \cite{Komargodski:2011vj,Fernandes:2021dsb}.
Under these transformations, the infinitesimal changes of 
$g_{\mu \nu}$ and $\phi$ are given, respectively, by 
$\delta_{\sigma} g_{\mu \nu}=2\sigma g_{\mu \nu}$ and 
$\delta_{\sigma}\phi=\sigma$. 
Under such infinitesimal Weyl transformations, the action 
${\cal S}_A[g,\phi]$, which depends on $g_{\mu \nu}$ 
and $\phi$, varies by the amount 
\be
\delta_{\sigma}{\cal S}_A=\int {\rm d}^4 x 
\left( 2g_{\mu \nu} \frac{\delta {\cal S}_A[g,\phi]}{\delta g_{\mu \nu}}
+\frac{\delta {\cal S}_A[g,\phi]}{\delta \phi} \right)\sigma
=\int {\rm d}^4 x \sqrt{-g} \left( -T_A+\frac{1}{\sqrt{-g}}
\frac{\delta {\cal S}_A[g,\phi]}{\delta \phi} \right)\sigma\,,
\ee
where $T_A$ is the trace given by 
\be
T_A=-\frac{2}{\sqrt{-g}} g_{\mu \nu}
\frac{\delta {\cal S}_A[g,\phi]}{\delta g_{\mu \nu}}\,.
\ee
The conformal invariance requires that $\delta_{\sigma}{\cal S}_A=0$ 
for $\sigma \neq 0$ and hence
\be
\frac{\delta {\cal S}_A[g,\phi]}{\delta \phi}
=\sqrt{-g}\,T_A=\sqrt{-g} \left( -\alpha {\cal G}+\beta W^2 
\right)\,.
\label{traeq}
\ee
One can express ${\cal G}$, $W^2$ in $T_A$ and 
$\sqrt{-g}$ by using corresponding quantities in the 
conformally transformed frame with the metric 
$\bar{g}_{\mu \nu}=e^{-2\phi}g_{\mu \nu}$. 
In particular, the conformal transformation of the GB 
trace anomaly gives rise to a derivative 
coupling with the Einstein tensor $\bar{G}^{\mu \nu}$ 
and nonlinear derivative terms like 
$8 (\bar{\nabla}_{\mu}\phi 
\bar{\nabla}^{\mu}\phi) \bar{\square}\phi$ \cite{Dabrowski:2008kx}, 
where we use an overbar for the quantities and derivatives 
in the frame with the metric $\bar{g}_{\mu \nu}$. 
Then, Eq.~(\ref{traeq}) yields 
\ba
\frac{\delta {\cal S}_A[\bar{g},\phi]}{\delta \phi}
&=& 
-\sqrt{-\bar{g}}\,\alpha [ \bar{\cal G} 
-8 \bar{R}^{\mu \nu}\bar{\nabla}_{\mu}\phi 
\bar{\nabla}_{\nu}\phi +8 \bar{G}^{\mu \nu} 
\bar{\nabla}_{\mu} \bar{\nabla}_{\nu}\phi
+8 (\bar{\nabla}_{\mu}\phi 
\bar{\nabla}^{\mu}\phi) \bar{\square}\phi 
-8 \bar{\nabla}_{\mu} \bar{\nabla}_{\nu}\phi
\bar{\nabla}^{\mu} \bar{\nabla}^{\nu}\phi
+8 (\bar{\square} \phi)^2 \nonumber \\
& &+16 \bar{\nabla}_{\mu}\phi \bar{\nabla}_{\nu}\phi 
\bar{\nabla}^{\mu} \bar{\nabla}^{\nu} \phi ]
+\sqrt{-\bar{g}}\beta \bar{W}^2\,.
\label{traeq2}
\ea
The action ${\cal S}_A$ satisfying the relation (\ref{traeq})
can be constructed by considering a straight line path 
$\phi (\eta)=\eta \phi$ with $0 \le \eta \le 1$, 
which connects the values $\phi(0)=0$ and 
$\phi(1)=\phi$ \cite{Fernandes:2021dsb}. 
The resulting action is given by 
\ba
{\cal S}_A[\bar{g},\phi]
&=& \int {\rm d}^4 x 
\frac{\delta {\cal S}_A[\bar{g},\phi]}{\delta \phi}
\frac{{\rm d}\phi(\eta)}{{\rm d}\eta}=
\int {\rm d}^4 x 
\int_0^1 {\rm d}\eta 
\frac{\delta {\cal S}_A[\bar{g},\phi]}{\delta \phi}
\biggr|_{\phi \to \eta \phi} \phi \nonumber \\
&=& -\alpha \int {\rm d}^4 x \sqrt{-\bar{g}} 
\left( \phi \bar{\cal G} -4\bar{G}^{\mu \nu} 
\bar{\nabla}_{\mu}\phi \bar{\nabla}_{\nu}\phi
+8\bar{X} \bar{\square}\phi-8 \bar{X}^2 \right)
+\beta \int {\rm d}^4 x \sqrt{-\bar{g}}\,\phi \bar{W}^2\,,
\label{SA}
\ea
where $\bar{X}=-(1/2)\bar{g}^{\mu \nu} 
\bar{\nabla}_{\mu} \phi \bar{\nabla}_{\nu} \phi$. 
We can choose other paths connecting two points 
$\phi(0)=0$ and $\phi(1)=\phi$, but the resulting 
action is equivalent to Eq.~(\ref{SA}) \cite{Soper:1976bb}.
The action (\ref{SA}) coincides with Eq.~(8) of Ref.~\cite{Riegert:1984kt} 
originally derived by Riegert, but now the metric tensor $\bar{g}_{\mu \nu}$ 
is not subject to the fixed determinant constraint.

The Einstein-Hilbert action in GR is expressed as 
\be
{\cal S}_{\rm GR}
= M^2 \int {\rm d}^4 x \sqrt{-g} R
=M^2 \int {\rm d}^4 x \sqrt{-\bar{g}}\,e^{2\phi} 
\left( \bar{R}+6 \bar{g}^{\mu \nu} \bar{\nabla}_{\mu} \phi 
\bar{\nabla}_{\nu} \phi \right)\,,
\label{SGR}
\ee
where the second equality holds up to boundary terms. 
As we already mentioned, GR supplemented by the trace 
anomaly action (\ref{SA}) is an inconsistent EFT.
We observe that the action (\ref{SGR}) expressed in terms of 
the metric $\bar{g}_{\mu \nu}$ contains an apparent  
kinetic term of the scalar field, but it has a negative 
kinetic energy.
This kinetic term can be eliminated by the field 
redefinition $\bar{g}_{\mu \nu} \to e^{-2\phi} g_{\mu \nu}$, 
but nonlinear scalar field derivatives survive in the action 
of ${\cal S}_A [g, \phi]$. Hence such a theory is plagued by 
the strong coupling problem. 

It is worth mentioning that the action same as (\ref{SA}) also appears 
as a result of the regularized Kaluza-Klein reduction of 
$D~(>4)$-dimensional Einstein-Gauss-Bonnet (EGB) theory on a 
$(D-4)$-dimensional maximally symmetric space with a vanishing spatial curvature. In this scenario, the $D$-dimensional metric can be written 
in the form ${\rm d}s_D^2={\rm d}s_4^2+e^{-2\phi}{\rm d}\sigma_{D-4}^2$, where ${\rm d} s_4^2$ and ${\rm d} \sigma_{D-4}^2$ 
are the line elements of 
4-dimensional spacetime and internal space, respectively. 
Here, the scalar field $\phi$ corresponds to the size of internal space, 
which only depends on the 4-dimensional coordinate. 
The $D$-dimensional action of EGB theory is given by 
\be
{\cal S}_{\rm EGB}=M_D^2 \int {\rm d}^D x \sqrt{-g_D}
\left( R_D+\hat{\alpha} {\cal G}_D \right)\,, 
\label{SEGB}
\ee
where the subscript $``D$'' represents $D$-dimensional quantities, 
and $\hat{\alpha}$ is the GB coupling constant. 
Performing the volume integral of (\ref{SEGB}) under the above 
metric ansatz, we can express ${\cal S}_{\rm EGB}$ in terms of the 4-dimensional curvature quantities $R$, $G^{\mu \nu}$, ${\cal G}$, 
and the scalar field $\phi$ and its derivatives. 
In this process, the integration constant is absorbed into $M_D$ to 
define the 4-dimensional reduced Planck mass 
$\Mpl=\sqrt{2}M$.
We add a counter term $-M^2 \int {\rm d}^4 x 
\sqrt{-g}\,\hat{\alpha}{\cal G}$ to the action (\ref{SEGB}) 
and rescale the coupling constant 
as $\hat{\alpha} \to \alpha/(D-4)$ \cite{Glavan:2019inb}. 
Taking the $D\to 4$ limit in the end, we obtain 
the reduced 4-dimensional action 
\be
{\cal S}_{\rm 4DEGB}=M^2 \int {\rm d}^4 x \sqrt{-g}
\left[ R-\alpha ( \phi{\cal G} -4G^{\mu \nu} 
\nabla_{\mu}\phi \nabla_{\nu}\phi
+8X \square \phi-8X^2 ) \right]\,. 
\label{S4DEGB}
\ee
The terms proportional to $\alpha$ in Eq.~(\ref{S4DEGB}) 
are exactly the same as those associated with the GB
trace anomaly in the action (\ref{SA}). 
The action (\ref{S4DEGB}) has an Einstein-Hilbert term, 
but there is no kinetic term of the scalar field $\phi$. 
Thus, 4DEGB gravity also suffers from the strong 
coupling problem, as recognized in Refs.~\cite{Bonifacio:2020vbk,Tsujikawa:2022lww,Kase:2023kvq}.

\subsection{Gabadadze's action}

To circumvent the strong coupling problem present in GR supplemented 
by the trace anomaly action (\ref{SA}), 
Gabadadze \cite{Gabadadze:2023quw} proposed the action 
\be
{\cal S}={\cal S}_{R \bar{R}}+{\cal S}_A\,,
\label{Sga}
\ee
where ${\cal S}_{R \bar{R}}$ and ${\cal S}_A$ are given, respectively, 
by Eqs.~(\ref{SRbR}) and (\ref{SA}), and the new mass scale $\bar{M}$ 
is assumed to be much smaller than $M$.
In terms of the metric $\bar{g}_{\mu \nu}=e^{-2\phi}g_{\mu \nu}$, 
the action (\ref{Sga}) can be expressed as
\be
{\cal S}= \int {\rm d}^4 x \sqrt{-\bar{g}} 
\left[ M^2 e^{2\phi} \left( \bar{R}-12\bar{X} \right)
-\bar{M}^2 \bar{R} 
-\alpha 
\left( \phi \bar{\cal G} -4\bar{G}^{\mu \nu} 
\bar{\nabla}_{\mu}\phi \bar{\nabla}_{\nu}\phi
+8\bar{X} \bar{\square}\phi-8 \bar{X}^2 \right)
+\beta \phi \bar{W}^2 
\right]\,.
\label{Sga2}
\ee
Note that GR with the trace anomaly action corresponds to 
the limit $\bar{M} \to 0$. 
In terms of the metric $g_{\mu \nu}$, 
the action ${\cal S}_{R \bar{R}}$ 
in Eq.~(\ref{SRbR2}) contains a kinetic term 
$-6 g^{\mu \nu} \nabla_{\mu}\varphi \nabla_{\nu}\varphi$ with 
a correct sign (i.e., no ghost) for the canonically normalized
scalar field $\varphi=\bar{M}e^{-\phi}$. 
The appearance of this correct sign is the result of introducing 
the action $-\bar{M}^2 \int {\rm d}^4 x \sqrt{-\bar{g}}\,
\bar{R}$.
The second and third terms of (\ref{SRbR2}) correspond to 
the action of a conformally invariant scalar field, whose invariance 
is broken by the Einstein-Hilbert term. 

Under the transformations 
$g_{\mu \nu} \to g_{\mu \nu}(1+\mu \varphi/M)^2/(1-\mu^2)$ and 
$\varphi \to \varphi+\mu M (1-\varphi^2/M^2)/(1+\mu \varphi/M)$, where 
$\mu$ is an arbitrary constant, the action (\ref{SRbR2}) 
is invariant \cite{Gabadadze:2023quw}. 
There is also an invariant combination of the metric tensor 
$\hat{g}_{\mu \nu}=g_{\mu \nu} (1-\varphi^2/M^2)$. 
If the matter fields in standard model of 
particle physics are coupled to gravity through the metric 
$\hat{g}_{\mu \nu}$, the corresponding actions 
in the matter sector are also invariant under such transformations.
Provided that $\bar{M} \ll M$, one has 
$\hat{g}_{\mu \nu} \simeq g_{\mu \nu}$ and 
hence the matter fields are approximately coupled to the 
metric tensor $g_{\mu \nu}$. 

The flat space expansion of (\ref{Sga2}) can be performed 
by substituting 
$g_{\mu\nu}=e^{2\phi} \bar{g}_{\mu \nu}=
\eta_{\mu \nu} (1-\varphi^2/M^2)^{-1}$, where 
$\eta_{\mu\nu}$ is the Minkowski metric.  
Introducing a canonically normalized field $\pi \equiv \bar{M}\phi$, 
the action (\ref{Sga2}) contains the derivative terms 
\be
{\cal S} \supset \int {\rm d}^4 x \left[ -6e^{-2\pi/\bar{M}} (\partial \pi)^2
+\alpha \left\{ \frac{4 (\partial \pi)^2 \square \pi}{\bar{M}^3} 
+\frac{2(\partial \pi)^4}{\bar{M}^4}
\right\}\right]\,,
\ee
where we used the notation
$(\partial \pi)^2=\partial_{\mu}\pi \partial^{\mu}\pi$ and 
the partial derivative $\partial_{\mu}=\partial/\partial x^{\mu}$.
Nonlinear derivative terms are suppressed below the scale $\bar{M}$ 
in comparison to the canonical kinetic term, while the theory is strongly 
coupled above the scale $\bar{M}$. 
Hence the theory given by the action (\ref{Sga2}) is an EFT 
with trace anomaly corrections valid below the scale $\bar{M}~(\ll M)$. 
Unlike the Riegert's theory, the metric 
$\bar{g}_{\mu \nu}=e^{-2\phi} g_{\mu \nu}$ is not subject to 
a fixed determinant constraint. 
Then, the Gabadadze's theory has two tensor propagating 
degrees of freedom besides one scalar mode $\phi$.

Varying the action (\ref{Sga2}) with respect to $\bar{g}_{\mu \nu}$, 
it follows that 
\be
\left( M^2 e^{2\phi} -\bar{M}^2 \right) \bar{G}_{\mu \nu}
=-M^2 e^{2\phi} \left[ 2 \bar{\nabla}_{\mu} \phi \bar{\nabla}_{\nu} \phi 
-2 \bar{\nabla}_{\mu} \bar{\nabla}_{\nu} \phi
+2 \bar{g}_{\mu \nu} \left( \bar{\square} \phi -\bar{X} \right) 
\right]-\alpha \bar{{\cal H}}_{\mu \nu}
-2\beta \phi \left( 2 \bar{\nabla}^{\rho} \bar{\nabla}^{\sigma}
+\bar{R}^{\rho \sigma} \right) \bar{C}_{\mu \rho \nu \sigma}\,,
\label{fieldeq}
\ee
where 
\ba
\bar{{\cal H}}_{\mu \nu} &\equiv&
-4\bar{X} \bar{G}_{\mu \nu}
+4 \left( \bar{\nabla}_{\alpha} \phi \bar{\nabla}_{\mu} \phi 
-\bar{\nabla}_{\alpha} \bar{\nabla}_{\mu} \phi \right)
\left( \bar{\nabla}^{\alpha}\phi \bar{\nabla}_{\nu} \phi 
-\bar{\nabla}^{\alpha} \bar{\nabla}_{\nu} \phi \right)
+4 \left( \bar{\nabla}_{\mu} \phi \bar{\nabla}_{\nu} \phi 
-\bar{\nabla}_{\nu} \bar{\nabla}_{\mu} \phi \right) \bar{\square} \phi 
\nonumber \\
&&+\bar{g}_{\mu \nu} \left[ 2(\bar{\square} \phi)^2-4\bar{X}^2
+2\bar{\nabla}_{\beta} \bar{\nabla}_{\alpha} \phi 
\left( 2\bar{\nabla}^{\alpha} \phi \bar{\nabla}^{\beta} \phi
-\bar{\nabla}^{\beta} \bar{\nabla}^{\alpha} \phi \right) \right]
+4 \bar{{\cal P}}_{\mu \alpha \nu \beta} \left( \bar{\nabla}^{\alpha}\phi 
\bar{\nabla}^{\beta}\phi-\bar{\nabla}^{\beta} 
\bar{\nabla}^{\alpha} \phi \right)\,,
\ea
with 
\be
\bar{{\cal P}}_{\mu \alpha \nu \beta} \equiv 
\bar{R}_{\mu \alpha \nu \beta}+\bar{g}_{\mu \beta} \bar{R}_{\alpha \nu}
-\bar{g}_{\mu \nu} \bar{R}_{\alpha \beta}
+\bar{g}_{\alpha \nu} \bar{R}_{\mu \beta}
-\bar{g}_{\alpha \beta} \bar{R}_{\mu \nu}
+\frac{1}{2} \left( \bar{g}_{\mu \nu} \bar{g}_{\alpha \beta} 
-\bar{g}_{\mu \beta} \bar{g}_{\alpha \nu} \right) \bar{R}\,.
\ee
We take the trace of Eq.~(\ref{fieldeq}) by exerting $\bar{g}^{\mu \nu}$ 
and exploit the relation $12 \bar{X}-6\bar{\square}\phi
=e^{2\phi}R-\bar{R}$.
On using the property $\bar{g}^{\mu \nu} 
\bar{\cal P}_{\mu \alpha \nu \beta}=-\bar{G}_{\alpha \beta}$, 
it follows that $\bar{g}^{\mu \nu} \bar{{\cal H}}_{\mu \nu}=(e^{4\phi}{\cal G}
-\bar{\cal G})/2$. We also note that the divergence of the Weyl tensor 
vanishes, such that $\bar{g}^{\mu \nu}\bar{C}_{\mu \rho \nu \sigma}=0$.
Then, the trace of Eq.~(\ref{fieldeq}) is expressed as 
\be
2\bar{M}^2 \bar{R}-\alpha \bar{\cal G}
=e^{4\phi} \left( 2M^2 R-\alpha {\cal G} \right)\,.
\label{RGre}
\ee
Varying the action ${\cal S}_{R \bar{R}}= \int {\rm d}^4 x \sqrt{-\bar{g}} 
\left[ M^2 e^{2\phi}(\bar{R}-12\bar{X} \right)
-\bar{M}^2 \bar{R}]$ with respect to $\phi$, we obtain 
\be
\frac{\delta {\cal S}_{R \bar{R}}}{\delta \phi}
=2\sqrt{-\bar{g}}\,M^2 e^{2\phi} \left(
\bar{R}+12 \bar{X}-6 \bar{\square} \phi 
\right)=2\sqrt{-g}M^2 R\,.
\ee
{}From Eq.~(\ref{traeq}), the variation of the trace anomaly action is 
given by $\delta {\cal S}_A/\delta \phi=
\sqrt{-g} \left( -\alpha {\cal G}+\beta W^2 \right)$. 
Then, varying the action ${\cal S}={\cal S}_{R\bar{R}}+{\cal S}_A$ 
with respect to $\phi$ leads to 
\be
2 M^2 R-\alpha {\cal G}+\beta W^2=0\,,
\label{trace1}
\ee
which is the scalar field equation in the frame with the metric $g_{\mu \nu}$.
{}From Eqs.~(\ref{RGre}) and (\ref{trace1}), we obtain 
\be
2\bar{M}^2 \bar{R}-\alpha \bar{\cal G}+\beta \bar{W}^2=0\,.
\label{trace2}
\ee
The spacetime geometry associated with the metric 
$\bar{g}_{\mu \nu}$ is modified by the GB and Weyl terms 
through the trace anomaly Eq.~(\ref{trace2}). 
While Eq.~(\ref{trace1}) contains the mass $M$ of order 
the Planck scale $\Mpl$, it is replaced by the new mass 
scale $\bar{M}$ in Eq.~(\ref{trace2}).

The Weyl curvature term in Eq.~(\ref{Sga2}) gives rise to the field 
equations of motion higher than second order. 
If the Weyl trace anomaly is absent, i.e., 
\be
\beta=0\,,
\ee
then the action (\ref{Sga2}) belongs to a subclass of 
Horndeski theories \cite{Horndeski} given by 
\ba
{\cal S}
&=&\int {\rm d}^4 x \sqrt{-\bar{g}}\,
\biggl[ \bar{G}_2-\bar{G}_{3} \bar{\square} \phi 
+\bar{G}_{4} R +\bar{G}_{4,\bar{X}} \left\{ (\bar{\square} \phi)^{2}
-(\bar{\nabla}_{\mu} \bar{\nabla}_{\nu} \phi)
(\bar{\nabla}^{\mu} \bar{\nabla}^{\nu} \phi) \right\}
\nonumber \\
&&
+\bar{G}_{5} \bar{G}^{\mu \nu} \bar{\nabla}_{\mu}
\bar{\nabla}_{\nu} \phi -\frac{1}{6} \bar{G}_{5,\bar{X}}
\left\{ (\bar{\square} \phi)^{3}-3(\bar{\square} \phi)\,
(\bar{\nabla}_{\mu} \bar{\nabla}_{\nu} \phi)
(\bar{\nabla}^{\mu} \bar{\nabla}^{\nu} \phi)
+2( \bar{\nabla}^{\mu} \bar{\nabla}_{\alpha} \phi)
( \bar{\nabla}^{\alpha} \bar{\nabla}_{\beta} \phi)
( \bar{\nabla}^{\beta} \bar{\nabla}_{\mu} \phi) \right\} \biggr]\,,
\label{action}
\ea
where $\bar{G}_{j,\bar{X}} \equiv \partial \bar{G}_j/\partial \bar{X}$ 
(with $j=4,5$), and 
\ba
& &
\bar{G}_2=-12 M^2 \bar{X} e^{2\phi}+8\alpha \bar{X}^2\,,\qquad 
\bar{G}_3=8\alpha \bar{X}\,,\nonumber \\
& & 
\bar{G}_4=M^2 e^{2\phi}-\bar{M}^2+4\alpha \bar{X}\,,\qquad 
\bar{G}_5=4\alpha \ln |\bar{X}| \,.
\label{G2345}
\ea
In particular, the linearly coupled GB Lagrangian 
$-\alpha \phi \bar{\cal G}$ can be accommodated 
by the quintic Horndeski function 
$\bar{G}_5=4\alpha \ln |\bar{X}|$ \cite{Kobayashi:2011nu,Langlois:2022eta}.
In this case, the field equations of motion are kept up to second order 
in both $\bar{g}_{\mu \nu}$ and $\phi$.

\section{Hairy black hole}
\label{BHsec}

We study the existence of hairy BH solutions for the action
(\ref{Sga2}) with $\beta=0$, i.e., 
\be
{\cal S}= \int {\rm d}^4 x \sqrt{-\bar{g}} 
\left[ \left( M^2 e^{2\phi}-\bar{M}^2 \right) \bar{R}
-12M^2 e^{2\phi}\bar{X} 
-\alpha 
\left( \phi \bar{\cal G} -4\bar{G}^{\mu \nu} 
\bar{\nabla}_{\mu}\phi \bar{\nabla}_{\nu}\phi
+8\bar{X} \bar{\square}\phi-8 \bar{X}^2 \right)
\right]\,,
\label{Sga3}
\ee
which is equivalent to the Horndeski action (\ref{action}) 
with the coupling functions (\ref{G2345}).
For completeness we need to take into account the Weyl 
trace anomaly term, but in this paper we would like to clarify 
whether or not the GB trace anomaly can induce linearly 
stable hairy BH solutions.
In terms of the metric tensor $\bar{g}_{\mu \nu}$, 
the static and spherically symmetric spacetime  
is given by the line element
\be 
\rd \bar{s}^2=-\bar{f}(\bar{r}) \rd t^{2} +\bar{h}^{-1}(\bar{r}) 
\rd \bar{r}^{2}+ \bar{r}^{2} \left( \rd \theta^{2}
+\sin^{2}\theta\,\rd\varphi^{2} 
\right)\,,
\label{background}
\ee
where the metric components $\bar{f}$ and $\bar{h}$ 
depend on the radial coordinate $\bar{r}$. 
The scalar field is assumed to be a function of 
$\bar{r}$ alone, i.e., $\phi=\phi(\bar{r})$.

We also write the line element associated with the metric 
$g_{\mu \nu}=e^{2\phi} \bar{g}_{\mu \nu}$ as 
\be 
\rd s^2=-f(r) \rd t^{2} +h^{-1}(r) 
\rd r^{2}+ r^{2} \left( \rd \theta^{2}
+\sin^{2}\theta\,\rd\varphi^{2} 
\right)\,,
\label{background2}
\ee
which is related to (\ref{background}) according 
to ${\rm d}\bar{s}^2=e^{-2\phi}{\rm d}s^2$. 
There are the following relations
\be
f=\bar{f} e^{2\phi}\,,\qquad 
h=\bar{h} \left[ 1+\bar{r} \phi'(\bar{r}) \right]^2\,,\qquad
r=\bar{r}e^{\phi}\,,
\label{fhr}
\ee
where a prime represents the derivative with respect to $\bar{r}$. 
In the following, we will obtain the solutions to $\bar{f}$, $\bar{h}$, 
and $\phi$ as functions of $\bar{r}$. 
Using the correspondence (\ref{fhr}), we will also derive
the functions $f(r)$ and $h(r)$ in the line element (\ref{background2}).

The differential equations for the metric components 
$\bar{f}(\bar{r})$ and $\bar{h}(\bar{r})$ are given by 
\ba
\frac{\baf'}{\baf} &=&
\frac{\bar{M}^2 (\bah-1)-M^2 e^{2\phi}
[\bah(3\bar{r}\phi'+1)(\bar{r}\phi'+1)-1]
-\alpha \bah \phi'^2 [ \bar{h}\{ \bar{r}\phi'(3\bar{r}\phi'+8)+6\}-2]}
{\bah [M^2 \bar{r} (\bar{r} \phi'+1)e^{2\phi}
+2\alpha \phi' (\bah \{3+\bar{r}\phi'(\bar{r}\phi'+3)\}-1)
-\bar{M}^2 \bar{r}]}\,,\label{baeq1} \\
\frac{\bah'}{\bah}-\frac{\baf'}{\baf}&=&
\frac{2(\phi'^2-\phi'')[M^2 \bar{r}^2 e^{2\phi}+2\alpha 
(\bar{r}^2 \bah \phi'^2+2\bar{r} \bah \phi'+\bah-1)]}
{\bar{r}M^2 e^{2\phi}(\bar{r} \phi'+1)-\bar{r} \bar{M}^2+2\alpha \phi' 
(\bar{r}^2 \bah \phi'^2+3\bar{r} \bah \phi'+3\bah-1)
}\,.\label{baeq2}
\ea
{}From Eq.~(\ref{trace2}), we obtain 
\ba
& &
\left[ 2\bar{r}^2 \baf \bah \baf''-\bar{r} \baf' \left( \bar{r} 
\bah \baf'-\bar{r} \baf \bah' -4\baf \bah \right)
+4\baf^2 (\bah-1) +4\bar{r} \baf^2 \bah' \right] \bar{M}^2 
\nonumber \\
& &
+2\alpha \left[ 2\baf \bah (\bah-1) \baf''
-\baf' (\bah^2 \baf'-3\baf \bah \bah'-\baf' \bah+\baf \bah') 
\right]=0\,.\label{baeq3}
\ea

In the absence of the GB trace anomaly, i.e., $\alpha=0$, 
there is the following exact solution to Eqs.~(\ref{baeq1})-(\ref{baeq3}):
\be
\bar{f}(\bar{r})=c_0 \left( 1-\frac{m_0}{\bar{r}} \right)^2\,,\qquad 
\bar{h}(\bar{r})=\left( 1-\frac{m_0}{\bar{r}} \right)^2\,,\qquad 
e^{\phi(\bar{r})}=\pm \frac{\bar{M}}{M}\frac{m_0}{\bar{r}-m_0}\,,
\label{Beken}
\ee
where $c_0$ and $m_0$ are constants. 
For a conformal scalar field with the Einstein-Hilbert action, 
the same type of solution was obtained 
by Bocharova, Bronnikov, and Melnikov \cite{Bocharova:1970skc} 
and by Bekenstein \cite{Bekenstein:1974sf} (BBMB).
However, it is unstable against monopole perturbations \cite{Bronnikov:1978mx,McFadden:2004ni,Konoplya:2005et} 
as well as perturbations for general multipoles $l$ \cite{Kobayashi:2014wsa}. 
This instability is also related to the divergence of 
$\phi'(\bar{r})=1/(m_0-\bar{r})$ on the horizon 
located at $\bar{r}=m_0$ \cite{Minamitsuji:2022mlv,Minamitsuji:2022vbi}.
In Appendix \ref{AppA}, we will show that the solution (\ref{Beken}) is unstable 
against linear perturbations for the radius $\bar{r}>2m_0$.

For $\alpha=0$, there is also the GR branch given by 
\be
\bar{f}(\bar{r})=\bar{h}(\bar{r})=1-\frac{\bar{r}_h}{\bar{r}}\,,\qquad 
\phi(\bar{r})=\phi_0={\rm constant}\,,
\label{Sch}
\ee
where $\bar{r}_h$ is the horizon radius. 
This solution is stable against linear perturbations. 
Thus, instead of (\ref{Beken}), the Schwarzschild branch 
without a scalar hair (\ref{Sch}) should be selected as 
a linearly stable BH.

The no-hair property of BHs for $\alpha=0$ changes in the 
presence of the GB trace anomaly. Let us derive the solutions 
to $\bar{f}$, $\bar{h}$, and $\phi$ under the expansion of 
a small coupling constant $\alpha$. 
In the limit that $\alpha \to 0$, the solutions need to recover the 
Schwarzschild metric without the scalar hair, i.e., Eq.~(\ref{Sch}).
Outside the horizon characterized by the radius $\bar{r}_h$, 
we search for solutions in the forms 
\be
f(\bar{r})=\left( 1-\frac{\bar{r}_h}{\bar{r}} \right) 
\left[ 1+ \sum_{i=1} \bar{f}_i(\bar{r}) \alpha^i \right]\,,\qquad
h(\bar{r})=\left( 1-\frac{\bar{r}_h}{\bar{r}} \right) 
\left[ 1+ \sum_{i=1} \bar{h}_i(\bar{r}) \alpha^i \right]\,,\qquad 
\phi(\bar{r})=\phi_0+\sum_{i=1}\phi_i (\bar{r}) \alpha^i\,,
\label{fhexpan}
\ee
where $\bar{f}_i$, $\bar{h}_i$, and $\phi_i$ are functions of $\bar{r}$. 
Substituting Eq.~(\ref{fhexpan}) into Eqs.~(\ref{baeq1})-(\ref{baeq3}), 
we obtain the differential equations for $\bar{f}_i$, $\bar{h}_i$, 
and $\phi_i$ at each order in $\alpha^i$.
We can also derive a second-order differential equation for $\phi(\bar{r})$ 
by differentiating Eq.~(\ref{baeq1}) with respect to $\bar{r}$ and 
eliminate the derivatives $\bar{f}'$, $\bar{h}'$, and $\bar{f}''$ by 
using Eqs.~(\ref{baeq1}), (\ref{baeq2}), and (\ref{baeq3}). 
At first order in the expansion of $\phi(\bar{r})$, we have 
\be
\phi_1''(\bar{r})+\frac{2\bar{r}-\bar{r}_h}{\bar{r}(\bar{r}-\bar{r}_h)}
\phi_1'(\bar{r})
-\frac{\bar{r}_h^2 (M^2-\bar{M}^2 e^{-2\phi_0})}
{M^2 \bar{M}^2 \bar{r}^5 (\bar{r}-\bar{r}_h)}=0\,.
\label{phi1}
\ee
The integrated solution to $\phi_1 (\bar{r})$ contains two integration constants. 
They are determined by imposing the regular boundary conditions 
$\phi_1'(\bar{r}_h)={\rm constant}$ on the horizon and 
$\phi_1(\infty) \to 0$ at spatial infinity. 
The finiteness of $\phi'(\bar{r})$ at $\bar{r}=\bar{r}_h$ is not only required 
for the validity of the expansion (\ref{fhexpan}) but also for 
avoiding the instability of linear perturbations \cite{Minamitsuji:2022mlv,Minamitsuji:2022vbi}. 
Then, the resulting integrated solution to Eq.~(\ref{phi1}) yields
\be
\phi_1 (\bar{r})=-\frac{(6\bar{r}^2+3\bar{r}_h \bar{r}
+2\bar{r}_h^2)(M^2-\bar{M}^2 e^{-2\phi_0})}
{18 M^2 \bar{M}^2 \bar{r}_h \bar{r}^3}\,.
\ee
Substituting this solution into Eq.~(\ref{baeq2}) and 
using Eq.~(\ref{baeq1}), we obtain the differential equation 
for $\bar{h}_1(\bar{r})$ as 
\be
\bar{h}_1'(\bar{r})+\frac{\bar{h}_1(\bar{r})}{\bar{r}-\bar{r}_h}
-\frac{\bar{r}^2+\bar{r}_h \bar{r}-5\bar{r}_h^2}
{3\bar{M^2}\bar{r}^4 (\bar{r}-\bar{r}_h)}=0\,.
\ee
The integrated solution respecting the finiteness of 
$\bar{h}_1(\bar{r})$ on the horizon is given by 
\be
\bar{h}_1(\bar{r})=-\frac{(\bar{r}+2\bar{r}_h)
(\bar{r}+5\bar{r}_h)}{18 \bar{M}^2 \bar{r}_h \bar{r}^3}\,, 
\ee
which approaches 0 at spatial infinity. 

{}From Eq.~(\ref{baeq1}), the differential equation for 
$\bar{f}_1(\bar{r})$ is
\be
\bar{f}_1'(\bar{r})+\frac{23 \bar{r}^2+22 \bar{r}_h \bar{r}
+18 \bar{r}_h^2}{18 \bar{M}^2 \bar{r}_h \bar{r}^4}=0\,.
\ee
The integrated solution satisfying the boundary condition 
$\bar{f}_1(\infty)=0$ is 
\be
\bar{f}_1 (\bar{r})=\frac{23\bar{r}^2+11\bar{r}_h \bar{r}+6\bar{r}_h^2}
{18 \bar{M}^2 \bar{r}_h \bar{r}^3}\,,
\ee
which decreases as $\bar{f}_1 (\bar{r}) \propto \bar{r}^{-1}$ 
at large distances. 

Similarly, we derive the solutions to $\bar{f}_i (\bar{r})$, $\bar{h}_i (\bar{r})$, 
and $\phi_i (\bar{r})$ at each order in $\alpha^i$ by 
imposing the regular boundary conditions explained above. 
As we will see in Sec.~\ref{stasec}, the expansion up to the 
order of $i=4$ is required for the purpose of studying the 
propagation of linear perturbations correctly.
Due to the complexity of functions $\bar{f}_i (\bar{r})$, $\bar{h}_i (\bar{r})$, 
$\phi_i (\bar{r})$ for $i \geq 3$, 
we only write the solutions up to the $i=2$ order as 
\ba
\bar{f}(\bar{r}) &=&
\left( 1 - \frac{\bar{r}_h}{\bar{r}} \right) 
\biggl[ 1+\frac{23\bar{r}^2+11\bar{r}_h \bar{r}+6\bar{r}_h^2}
{18 \bar{M}^2 \bar{r}_h \bar{r}^3}\alpha
-\frac{1}{32400 M^2 \bar{M}^4 \bar{r}_h^3 \bar{r}^6}
\{  \bar{M}^2 (38731\bar{r}^5+26371 \bar{r}_h \bar{r}^4
+22721 \bar{r}_h^2 \bar{r}^3 \nonumber \\
& & -769 \bar{r}_h^3 \bar{r}^2-5572  
\bar{r}_h^4 \bar{r}-9300 \bar{r}_h^5 ) e^{-2\phi_0}
-M^2 (6979 \bar{r}^5+26539 \bar{r}_h \bar{r}^4+28489 
\bar{r}_h^2 \bar{r}^3+33379 \bar{r}_h^3 \bar{r}^2
\nonumber \\
& & 
+13492 \bar{r}_h^4 \bar{r}+4500 \bar{r}_h^5 ) \} \alpha^2
+{\cal O} (\alpha^3)
\biggr]\,,\label{fso} \\
\bar{h}(\bar{r}) &=&
\left( 1 - \frac{\bar{r}_h}{\bar{r}} \right) 
\biggl[ 1-\frac{(\bar{r}+2\bar{r}_h)
(\bar{r}+5\bar{r}_h)}{18 \bar{M}^2 \bar{r}_h \bar{r}^3}\alpha
-\frac{1}{32400 M^2 \bar{M}^4 \bar{r}_h^3 \bar{r}^6}
\{  \bar{M}^2 (21211\bar{r}^5+13231 \bar{r}_h \bar{r}^4
+11041\bar{r}_h^2 \bar{r}^3 \nonumber \\
& & -33019 \bar{r}_h^3 \bar{r}^2-36964 
\bar{r}_h^4 \bar{r}-40300 \bar{r}_h^5 ) e^{-2\phi_0}
+M^2 (\bar{r}-\bar{r}_h) \nonumber \\
& & \times 
(1901 \bar{r}^4-4178 \bar{r}_h \bar{r}^3
-15747 \bar{r}_h^2 \bar{r}^2-4976\bar{r}_h^3 \bar{r}
-1300 \bar{r}_h^4) \} \alpha^2
+{\cal O} (\alpha^3)
\biggr]\,,\label{hso} \\
\phi(\bar{r}) &=& \phi_0-\frac{(6\bar{r}^2+3\bar{r}_h \bar{r}
+2\bar{r}_h^2)(M^2-\bar{M}^2 e^{-2\phi_0})}
{18 M^2 \bar{M}^2 \bar{r}_h \bar{r}^3}\alpha
-\frac{1}{32400 M^4 \bar{M}^4 \bar{r}_h^3 \bar{r}^6}
[ 3M^4 (740 \bar{r}^5-130 \bar{r}_h \bar{r}^4-20 \bar{r}_h^2 
\bar{r}^3 \nonumber \\
& &
+1285 \bar{r}_h^3 \bar{r}^2+688 \bar{r}_h^4 \bar{r}
+300 \bar{r}_h^5) 
-200M^2\bar{M}^2 e^{-2\phi_0} (33 \bar{r}^5+18 \bar{r}_h \bar{r}^4
+16 \bar{r}_h^2 \bar{r}^3+33 \bar{r}_h^3 \bar{r}^2
+15 \bar{r}_h^4 \bar{r}+6 \bar{r}_h^5)\nonumber \\
& &
+\bar{M}^4 e^{-4\phi_0} (4380 \bar{r}^5+3990 \bar{r}_h 
\bar{r}^4+3260 \bar{r}_h^2 \bar{r}^3+2745 \bar{r}_h^3 \bar{r}^2
+936 \bar{r}_h^4 \bar{r}+300\bar{r}_h^5)]\alpha^2
+{\cal O} (\alpha^3)\,.\label{phiso}
\ea
Outside the horizon, the leading-order trace anomaly corrections to 
$\bar{f} (\bar{r})$ and $\bar{h} (\bar{r})$ are largest around 
$\bar{r}=\bar{r}_h$, which are of order $\alpha/(\bar{M}^2 \bar{r}_h^2)$. 
For the validity of the expansion (\ref{fhexpan}), we then require that 
\be
\hat{\alpha} \equiv \frac{\alpha}
{\bar{M}^2 \bar{r}_h^2} \ll 1\,.
\label{alcon}
\ee
Provided that $\bar{M} \ll M$, the second-order 
corrections to Eqs.~(\ref{fso}) and (\ref{hso}) are at most 
of order $\hat{\alpha}^2$.
{}From Eq.~(\ref{phiso}), the first- and second-order 
corrections to $\phi(\bar{r})$ are at most of orders 
$\hat{\alpha}$ and $\hat{\alpha}^2$, respectively.
In the limit that $\bar{M} \to 0$, the condition (\ref{alcon}) is violated 
and hence the expanded solutions (\ref{fso})-(\ref{phiso}) are invalid 
in GR supplemented by the trace 
anomaly action ${\cal S}_A$. 
For the mass scale $\bar{M}$ larger than 
$1/\bar{r}_h$, the inequality (\ref{alcon}) is satisfied for $\alpha \ll 1$. 
This means that $\bar{M}$ can be chosen down to the order 
$1/\bar{r}_h$. For $\bar{M} \simeq 1/\bar{r}_h$, 
the EFT is valid for the length scale larger than $\bar{r}_h$.

Unlike the BBMB solution, the field derivative $\phi'(\bar{r})$ 
is finite on the horizon. 
At large distances, the scalar field solution up 
to the order of $\alpha$ is given by 
\be
\phi(\bar{r})=\phi_0-\frac{q_s}{\bar{r}}\qquad {\rm for} \qquad 
\bar{r} \gg \bar{r}_s\,,
\label{phiqs}
\ee
where $q_s$ is regarded as a scalar charge given by 
\be
q_s=\frac{M^2-\bar{M}^2e^{-2\phi_0}}
{3M^2 \bar{M}^2 \bar{r}_h}\alpha\,.
\label{qs}
\ee
The GB trace anomaly gives rise to a hairy BH solution 
possessing the scalar charge $q_s$.

\begin{figure}[ht]
\begin{center}
\includegraphics[height=3.2in,width=3.2in]{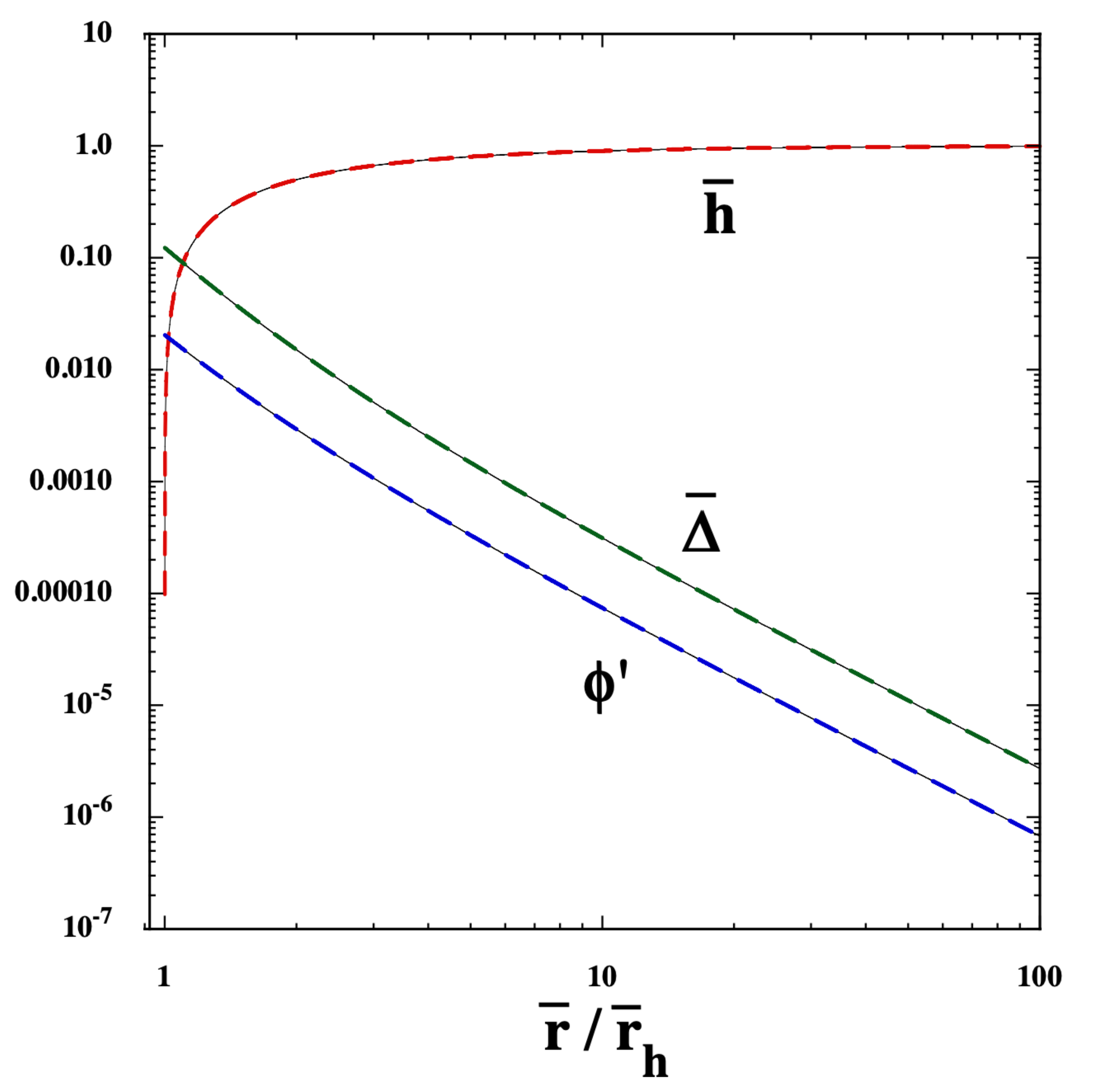}
\end{center}\vspace{-0.5cm}
\caption{\label{fig1}
We plot $\bar{h}$, $\bar{\Delta}=\bar{r}_h (\bar{h}'/\bar{h}-\bar{f}'/\bar{f})$, 
and $\phi'$ versus $\bar{r}/\bar{r}_h$ for $\hat{\alpha}=0.02$ and 
$\bar{M}=10^{-38}M$ (black lines). 
The numerical integration is performed outward 
by using Eqs.~(\ref{fso})-(\ref{phiso}) as boundary 
conditions at $\bar{r}/\bar{r}_h=1.0001$ with $\phi_0=0$.
The colored dashes lines are those derived by using the 
solutions (\ref{fso})-(\ref{phiso}) expanded up to fourth order 
in $\alpha^i$, which agree well with the numerical results. }
\end{figure}

{}From Eqs.~(\ref{fso}) and (\ref{hso})
we find that $\bar{f}(\bar{r})$ and $\bar{h}(\bar{r})$ are not identical to 
each other for $\alpha \neq 0$.
This is different from an exact BH solution obtained by assuming 
$\bar{f}(\bar{r})=h(\bar{r})$ in a similar conformally 
invariant theory \cite{Fernandes:2021dsb}. 
In Appendix \ref{AppB}, we will derive such a solution by imposing 
the condition $\bar{f}(\bar{r})=h(\bar{r})$ in Eq.~(\ref{baeq3}).
This condition demands that the right hand side of Eq.~(\ref{baeq2}) 
vanishes. However, the scalar field solution derived in this way 
is not consistent with the regular boundary condition on the horizon 
as well as the other background equations like Eq.~(\ref{baeq1}). 
In other words, the solutions satisfying all the equations of motion and 
regular boundary conditions on the horizon are of the forms 
(\ref{fso})-(\ref{phiso}). They are also different from the hairy BH 
solution present for a canonical scalar field $\phi$ linearly coupled to 
the GB term \cite{Sotiriou:2013qea,Sotiriou:2014pfa} 
(see also Refs.~\cite{Kanti:1995vq,Kanti:1997br,Torii:1996yi,Chen:2006ge,Guo:2008hf,Pani:2009wy,Cai:2009ua,Kleihaus:2011tg,Ayzenberg:2014aka,Cai:2014jea,Maselli:2015tta,Kleihaus:2015aje,Minamitsuji:2022mlv,Minamitsuji:2022vbi} 
for related works). 
In the latter theory the first-order metric components 
$\bar{f}_1$ and $\bar{h}_1$ vanish \cite{Sotiriou:2014pfa,Minamitsuji:2022mlv}, 
but this is not the case 
for our theory due to the existence of additional nonlinear
derivative terms in Eq.~(\ref{Sga3}). 

In Fig.~\ref{fig1}, we show numerically integrated solutions to 
$\bar{h}$, $\bar{\Delta} \equiv \bar{r}_h (\bar{h}'/\bar{h}-\bar{f}'/\bar{f})$, 
and $\phi'$ for $\hat{\alpha}=0.02$ and $\bar{M}=10^{-38}M$. 
This choice of $\bar{M}$ corresponds to the length scale of order 
$1/\bar{M} \simeq 10$~km. 
Provided that $\bar{M} \ll M$, the leading-order trace anomaly 
corrections to Eqs.~(\ref{fso})-(\ref{phiso}) are determined by 
the coupling $\hat{\alpha}=\alpha/(\bar{M}^2 \bar{r}_h^2)$.
For $\bar{M} \simeq 1/\bar{r}_h$, the corrections to 
$\bar{f}(\bar{r})$, $\bar{h}(\bar{r})$, and $\phi (\bar{r})$ 
are at most of order $\alpha$.
In Fig.~\ref{fig1}, we observe that the numerical results are in good agreement 
with the analytic solutions (\ref{fso})-(\ref{phiso}) expanded up to fourth
order in $\alpha^i$. The field derivative $\phi'$, which is finite on the horizon, decreases as
$\phi'(\bar{r}) \propto \bar{r}^{-2}$ in the regime $\bar{r} \gg \bar{r}_h$. 
Outward from the horizon, $\bar{h}$ continues to increase 
toward the asymptotic value 1. The quantity $\bar{\Delta}$, 
which characterizes the difference between 
$\bar{h}$ and $\bar{f}$, is largest around $\bar{r}=\bar{r}_h$. 
{}From Eqs.~(\ref{fso}) and (\ref{hso}), we obtain the asymptotic 
behavior $\bar{\Delta} \simeq 4\alpha/(3\bar{M}^2 \bar{r}^2)$ 
in the regime $\bar{r} \gg \bar{r}_h$, whose dependence 
can be also confirmed in Fig.~\ref{fig1}. 

The metric components $f$ and $h$ in the frame with 
the metric tensor $g_{\mu \nu}$ can be obtained as functions of $r$ 
by using the correspondence (\ref{fhr}). 
Taking $\phi_0=0$, the resulting forms of $f(r)$ and $h(r)$ correspond to 
those derived by the replacements $\bar{r} \to r$, $\bar{r}_h \to r_h$, 
$\bar{M} \to M$, and $M \to \bar{M}$ in Eqs.~(\ref{fso}) 
and (\ref{hso}), i.e.,
\ba
f(r) &=&
\left( 1 - \frac{r_h}{r} \right) 
\biggl[ 1+\frac{23r^2+11r_h r+6 r_h^2}
{18 M^2 r_h r^3}\alpha
-\frac{1}{32400 \bar{M}^2 M^4 r_h^3 r^6}
\{ M^2 (38731 r^5+26371 r_h r^4
+22721 r_h^2 r^3 \nonumber \\
& & -769 r_h^3 r^2-5572  
r_h^4r-9300 r_h^5 ) 
-\bar{M}^2 (6979 r^5+26539 r_h r^4+28489 
r_h^2 r^3+33379 r_h^3 r^2
\nonumber \\
& & 
+13492 r_h^4 r+4500 r_h^5 ) \} \alpha^2
+{\cal O} (\alpha^3)
\biggr]\,,\label{fsod} \\
h(r) &=&
\left( 1 - \frac{r_h}{r} \right) 
\biggl[ 1-\frac{(r+2r_h)
(r+5r_h)}{18 M^2 r_h r^3}\alpha
-\frac{1}{32400 \bar{M}^2 M^4 r_h^3 r^6}
\{ M^2 (21211 r^5+13231r_h r^4
+11041 r_h^2 r^3 \nonumber \\
& & -33019 r_h^3 r^2-36964 
r_h^4 r-40300 r_h^5 ) 
+\bar{M}^2 (r-r_h) \nonumber \\
& & \times 
(1901r^4-4178 r_h r^3
-15747 r_h^2 r^2-4976 r_h^3 r
-1300 r_h^4) \} \alpha^2
+{\cal O} (\alpha^3)
\biggr]\,.
\label{hsod} 
\ea
The leading-order trace anomaly corrections to $f(r)$ and $h(r)$ 
are subject to strong suppression by the appearance of 
the mass $M=\Mpl/\sqrt{2}$ in their denominators. 
As we already mentioned, the matter fields feel the gravitational 
force through the metric tensor $\hat{g}_{\mu \nu} \simeq g_{\mu \nu}$. 
Let us consider a test particle with mass $m_g$ on the metric 
background $g_{\mu \nu}$. 
Defining the gravitational potential $\Psi(r)$ as $f(r)=1+2\Psi(r)$ 
in the regime away from the horizon, 
the force exerting on the particle is given by 
\be
F(r)=-m_g \frac{{\rm d}\Psi (r)}{{\rm d}r}
=-\frac{m_g}{2} \frac{{\rm d}f(r)}{{\rm d}r}
=-\frac{m_g r_h}{2r^2} \left[ 
1- \left( \frac{23}{18}-\frac{4r_h}{3r}-\frac{5r_h^2}{6r^2} 
-\frac{4r_h^3}{3r^3} \right) \frac{\alpha}{M^2 r_h^2}
+{\cal O}(\alpha^2) \right]\,,
\label{Fg}
\ee
where we used the expanded solution (\ref{fsod}), and 
$r$ is related to $\bar{r}$ according to 
\be
r=\bar{r} \left[ 1-\left( 1-\frac{\bar{M}^2}{M^2} 
\right) \left( \frac{\bar{r}_h}{3\bar{r}}+
\frac{\bar{r}_h^2}{6\bar{r}^2}+
\frac{\bar{r}_h^3}{9\bar{r}^3} \right) 
\hat{\alpha }+{\cal O} ( \hat{\alpha}^2) \right]\,.
\label{rbr}
\ee
In terms of the distance $\bar{r}$ and horizon radius $\bar{r}_h$, 
the force (\ref{Fg}) can be expressed as
\ba
F(r) &=& 
-\frac{m_g e^{\phi}}{2} \frac{\bar{f}'+2\bar{f}\phi'}
{1+\bar{r}\phi'} \nonumber \\
&=& -\frac{m_g \bar{r}_h}{2\bar{r}^2} 
\left[ 1- \left\{ \frac{11}{18}-\frac{2\bar{r}_h}{3\bar{r}}
-\frac{\bar{r}_h^2}{3\bar{r}^2} 
-\frac{2\bar{r}_h^3}{9 \bar{r}^3}
+\frac{\bar{M}^2}{M^2}
\left( \frac{2}{3}-\frac{2\bar{r}_h}{3\bar{r}}
-\frac{\bar{r}_h^2}{2\bar{r}^2}
-\frac{10\bar{r}_h^3}{9\bar{r}^3} \right) \right\} \hat{\alpha}
+{\cal O}(\hat{\alpha}^2) \right]\,.
\label{Frd}
\ea
The $\alpha$-dependent terms in Eq.~(\ref{Fg}) correspond to 
fifth-force corrections to the gravitational force $F_g=-m_g r_h/(2r^2)$.
The leading-order fifth force relative to $F_g$ is at most of order 
$(\bar{M}/M)^2 \hat{\alpha}$, so it is even more suppressed than 
$\hat{\alpha}$ due to the small ratio $(\bar{M}/M)^2 \ll 1$. 
In Eq.~(\ref{Frd}), on the other hand, the leading-order trace anomaly 
correction relative to  $-m_g \bar{r}_h/(2\bar{r}^2)$ is of order $\hat{\alpha}$. 
This difference arises from the fact that the relative difference between 
$r$ and $\bar{r}$ is at most of order $\hat{\alpha}$, 
see Eq.~(\ref{rbr}).
When we express $r$ in terms of $\bar{r}$, the order $\hat{\alpha}$ 
correction, which is much larger than $(\bar{M}/M)^2 \hat{\alpha}$, 
appears in Eq.~(\ref{Frd}). 
Since the test particle feels gravity associated with 
the metric $g_{\mu \nu}$ and distance $r$, 
we need to interpret that the fifth force is suppressed by 
the factor $(\bar{M}/M)^2 \hat{\alpha}$ relative to the gravitational 
force $F_g=-m_g r_h/(2r^2)$. 
If we consider astrophysical BHs whose horizon radii $r_h$ 
are larger than the order 10~km, then the ratio $(\bar{M}/M)^2$
is significantly smaller than 1 for $\bar{M}$ of order
$1/r_h$. This is not necessarily the case for microscopic BHs with smaller 
horizon sizes, in which case the ratio $(\bar{M}/M)^2$ can be larger.

While the fifth force exerting on the test particle can be suppressed for 
astrophysical BHs, the scalar field (\ref{phiso}) receives the trace anomaly 
correction at most of order $\hat{\alpha}$ (even if we express $\phi$ 
with respect to $r$). During the inspiral phase of 
a BH binary system, the scalar radiation arising from the perturbation of 
$\phi$ may leave some signatures of the scalar charge 
in observed gravitational waveforms. 
This will deserve for a further detailed study.

\section{Linear stability of hairy black hole}
\label{stasec}

Finally, we study the linear stability of hairy BHs derived 
in Sec.~\ref{BHsec}. For this purpose, we exploit the solutions 
(\ref{fso})-(\ref{phiso}) obtained under the expansion 
of a small coupling constant $\alpha$ in the frame with 
the metric $\bar{g}_{\mu \nu}$. 
The perturbations on the static and spherically symmetric background 
can be decomposed into odd- and even-parity 
modes \cite{Regge:1957td,Zerilli:1970se}. 
In Refs.~\cite{Kobayashi:2012kh,Kobayashi:2014wsa}, the stability 
conditions of BHs against odd- and even-parity linear 
perturbations were derived except for the angular stability of 
even-parity modes.
In Ref.~\cite{Kase:2021mix}, the authors incorporated a perfect fluid 
with the background density $\rho$ and pressure $P$ and obtained
all the linear stability conditions applicable to NSs 
as well (see also Ref.~\cite{Kase:2020qvz}).
We will exploit those results in the following discussion.

In Horndeski theories given by the action (\ref{Sga3}), the 
odd-parity sector has a gravitational wave mode $\chi$  
arising from the metric perturbation. 
For this perturbation $\chi$, there are neither ghost nor Laplacian 
instabilities under the following conditions
\ba
\bar{{\cal G}} &\equiv& 2\bar{G}_4+2 \bar{h} \phi'^2 \bar{G}_{4,\bar{X}}
-\bar{h} \phi'^2 \left( \bar{G}_{5,\phi}
+{\frac {\bar{f}' \bar{h} \phi' \bar{G}_{5,\bar{X}}}{2\bar{f}}} 
\right)=2(M^2 e^{2\phi}-\bar{M}^2)+4\alpha \frac{\bar{h}}{\bar{f}} 
\phi' (\bar{f} \phi'+\bar{f}')
>0\,,
\label{cGdef}\\
\bar{{\cal F}} &\equiv&2 \bar{G}_4+\bar{h}\phi'^2
\bar{G}_{5,\phi}-\bar{h}\phi'^2  \left( \frac12 \bar{h}' \phi'
+\bar{h} \phi'' \right) \bar{G}_{5,\bar{X}}
=2(M^2 e^{2\phi}-\bar{M}^2)+4\alpha (2\bar{h}\phi''
+\bar{h}' \phi'-\bar{h} \phi'^2)
>0\,,
\label{cFdef}\\
\bar{{\cal H}} &\equiv&2 \bar{G}_4+2 \bar{h}\phi'^2 
\bar{G}_{4,\bar{X}}-\bar{h}\phi'^2 \bar{G}_{5,\phi}
-\frac{\bar{h}^2 \phi'^3 \bar{G}_{5,\bar{X}}}{\bar{r}}
=2(M^2 e^{2\phi}-\bar{M}^2)+4\alpha 
\frac{\bar{h}}{\bar{r}} \phi' ( \bar{r}\phi'+2) >0
\,.\label{cHdef}
\ea
Substituting the solutions (\ref{fso})-(\ref{phiso}) into 
Eq.~(\ref{cGdef}), the no ghost parameter 
in the odd-parity sector yields
\be
\bar{{\cal G}}=2(M^2 e^{2\phi}-\bar{M}^2) 
\left[ 1-\frac{6 \bar{r}^2+3\bar{r}_h \bar{r}
+2\bar{r}_h^2}{9 \bar{M}^2 \bar{r}_h \bar{r}^3} 
\alpha+{\cal O}(\alpha^2) \right]\,.
\ee
Provided that $\hat{\alpha} \ll 1$, the no-ghost condition  
$\bar{{\cal G}}>0$ is satisfied if
\be
M^2 e^{2\phi}-\bar{M}^2>0\,,
\label{noghost}
\ee
which automatically holds for $\bar{M} \ll M$.
Up to linear order in $\alpha^i$, the expressions of $\bar{{\cal F}}$ 
and $\bar{{\cal H}}$ coincide with $\bar{{\cal G}}$. 
However, the difference appears at the order of $\alpha^2$. 
The squared propagation speeds of odd-parity gravitational 
perturbation $\chi$ along the radial and angular directions are 
given, respectively, by 
\ba
& &
\bar{c}_{r,{\rm odd}}^2=\frac{\bar{{\cal G}}}{\bar{{\cal F}}}
=1+\frac{4(\bar{r}-\bar{r}_h) (2\bar{r}^2+3 \bar{r}_h \bar{r}
+4\bar{r}_h^2)e^{-2\phi_0}}{3M^2 \bar{M}^2 \bar{r}_h \bar{r}^6}
\alpha^2+{\cal O}(\alpha^3)\,,\label{crodd} \\
& &
\bar{c}_{\Omega,{\rm odd}}^2=\frac{\bar{\cal G}}{\bar{\cal H}}
=1-\frac{2(2\bar{r}-3\bar{r}_h) (\bar{r}^2+\bar{r}_h \bar{r}
+\bar{r}_h^2)e^{-2\phi_0}}{3M^2 \bar{M}^2 \bar{r}_h \bar{r}^6}
\alpha^2+{\cal O}(\alpha^3)\,.\label{cOodd}
\ea
Since the trace anomaly corrections in Eqs.~(\ref{crodd}) and (\ref{cOodd}) 
are at most of order $(\bar{M}^2/M^2)\hat{\alpha}^2$, 
both $\bar{c}_{r,{\rm odd}}^2$ and $\bar{c}_{\Omega,{\rm odd}}^2$ 
are very close to 1 in the vicinity of the horizon.
Thus, the Laplacian stability conditions $\bar{c}_{r,{\rm odd}}^2>0$ 
and $\bar{c}_{\Omega,{\rm odd}}^2>0$ are automatically satisfied 
for $\hat{\alpha} \ll 1$.
We also note that the $\alpha^2$-order corrections 
to Eqs.~(\ref{crodd}) and (\ref{cOodd}) decrease 
in proportion to $\bar{r}^{-3}$ at large distances, so 
$\bar{c}_{r,{\rm odd}}^2$ and $\bar{c}_{\Omega,{\rm odd}}^2$
rapidly approach 1 far away from the horizon. 

In the even-parity sector, the ghost does not appear under 
the condition 
\be
\bar{{\cal K}} \equiv 2 \bar{{\cal P}}_1-\bar{{\cal F}}>0\,,
\label{K}
\ee
where 
\ba
\bar{{\cal P}}_1 &\equiv& 
\frac{\bar{h} \bar{\mu}}{2\bar{f} \bar{r}^2 \bar{{\cal H}}^2} 
\left( \frac{\bar{f} \bar{r}^4 \bar{{\cal H}}^4}{\bar{\mu}^2 \bar{h}} 
\right)' \,,\\ 
\bar{\mu} &\equiv& 
4\bar{r} \bar{G}_4+2 \bar{r}^2 \phi' \bar{G}_{4,\phi}
+2\bar{h}\bar{r} \phi'^2 (4\bar{G}_{4,\bar{X}}-3 \bar{G}_{5,\phi})
+\bar{h}\phi'^3 [ (1-5\bar{h}) \bar{G}_{5,\bar{X}} 
+\bar{r}^2(\bar{G}_{3,\bar{X}} -2\bar{G}_{4,\phi \bar{X}}) 
\nonumber \\
& & +\bar{h}\phi' (\bar{h}\phi' \bar{G}_{5,\bar{X}\bar{X}} 
- 4  \bar{r} G_{4,\bar{X}\bar{X}}+ 2\bar{r} \bar{G}_{5,\phi\bar{X}})]\,.
\label{defP1}
\ea
Using the expanded solutions (\ref{fso})-(\ref{phiso}), 
it follows that 
\be
\bar{{\cal K}}=\frac{2(M^2e^{2\phi_0}-\bar{M}^2)
(\bar{r}^2+\bar{r}_h \bar{r}+\bar{r}_h^2)^2e^{-2\phi_0}}
{3M^2 \bar{M}^2 \bar{r}_h^2 \bar{r}^6}\alpha^2
+{\cal O}(\alpha^3)\,.
\label{calK}
\ee
Under the condition (\ref{noghost}), the leading-order term 
of $\bar{{\cal K}}$ is positive and hence the 
ghost is absent in the even-parity sector as well.
We also note that the trace anomaly correction generates 
the nonvanishing kinetic term $\bar{{\cal K}}$, in which case 
the strong coupling problem is absent.

For even-parity modes, there are two perturbations $\psi$ and 
$\delta \phi$ arising from the gravitational and scalar field 
sectors, respectively \cite{Kobayashi:2012kh,Kobayashi:2014wsa,Kase:2021mix}. 
The radial propagation speed squared of $\psi$ is 
identical to $\bar{c}_{r,{\rm odd}}^2
=\bar{{\cal G}}/\bar{{\cal F}}$ in the odd-parity sector. 
The other radial propagation speed squared $\bar{c}_{r,\delta \phi}^2$
can be obtained by setting $\rho=0=P$ in Eq.~(5.30) of 
Ref.~\cite{Kase:2021mix}. 
On using Eqs.~(\ref{fso})-(\ref{phiso}), we obtain 
\be
\bar{c}_{r,\delta \phi}^2=1-\frac{4(2\bar{r}^2+3\bar{r}_h \bar{r}
+4\bar{r}_h^2)(M^4+M^2 \bar{M}^2 e^{-2\phi_0}
+\bar{M}^4 e^{-4\phi_0})(\bar{r}-\bar{r}_h)}
{9 M^4 \bar{M}^4 \bar{r}_h \bar{r}^6}\alpha^2
+{\cal O}(\alpha^3)\,.
\label{cdelta}
\ee
To derive this expression, we need to resort to the background 
solutions of $\bar{f}(\bar{r})$, $\bar{h}(\bar{r})$, and 
$\phi(\bar{r})$ expanded up to fourth order in $\alpha^i$. 
In other words, the third order expansion leads to 
a result different from Eq.~(\ref{cdelta}), but using the 
solutions higher than fourth order gives the same expression 
as Eq.~(\ref{cdelta}). The trace anomaly correction in 
$\bar{c}_{r,\delta \phi}^2$ is at most of order $\hat{\alpha}^2$ 
and it decreases in proportion to $\bar{r}^{-3}$ at large distances. 
Provided that $\hat{\alpha} \ll 1$, the radial Laplacian stability 
of $\delta \phi$ is always ensured.
 
The squared propagation speeds 
of $\psi$ and $\delta \phi$ in the angular direction are 
$\bar{c}_{\Omega \pm,{\rm even}}^2=-B_1 \pm \sqrt{B_1^2-B_2}$, 
where the explicit expressions of $B_1$ and $B_2$ are given, 
respectively, by Eqs.~(5.37) and (5.38) of Ref.~\cite{Kase:2021mix}.
For our hairy BH solution, we have 
\be
\bar{c}_{\Omega \pm,{\rm even}}^2
=1 \pm \frac{2\sqrt{3}\,\bar{r}_h e^{-\phi_0}}
{M \bar{M} \bar{r}^3}\alpha+{\cal O}(\alpha^2)\,.
\label{cOmees}
\ee
Hence the leading-order trace anomaly correction 
in $\bar{c}_{\Omega \pm,{\rm even}}^2$
is suppressed by the order $(\bar{M}/M)\hat{\alpha}$ 
and it decreases in proportion to $\bar{r}^{-3}$ for increasing $\bar{r}$. 

We have thus shown that, under the condition 
$M^2 e^{2\phi}-\bar{M}^2>0$, our hairy BH solution 
arising from the trace anomaly satisfies 
all the linear stability conditions for $\hat{\alpha} \ll 1$ 
and $\bar{M} \ll M$. 
All the propagation speeds discussed above are close to 1
with small corrections induced by the GB trace anomaly.

\section{Conclusions}
\label{consec}

In a new gravitational theory with the trace anomaly  
recently proposed by Gabadadze, 
we studied the existence of hairy BH solutions 
on the static and spherically symmetric background. 
Introducing the action $-\bar{M}^2 \int \rd^4 x \sqrt{-\bar{g}}\bar{R}$
besides the Einstein-Hilbert term allows a possibility for avoiding the 
strong coupling problem present in GR supplemented by the trace anomaly. 
In terms of the metric tensor $\bar{g}_{\mu \nu}=e^{-2\phi}g_{\mu \nu}$ 
invariant under Weyl transformations 
$g_{\mu \nu} \to e^{2\sigma} g_{\mu \nu}$ and 
$\phi \to \phi+\sigma$, the total action 
${\cal S}={\cal S}_{R \bar{R}}+{\cal S}_A$ is expressed 
in the form (\ref{Sga2}). 
This is the EFT valid below the mass scale $\bar{M}$, 
in which regime nonlinear derivatives of the conformal 
scalar field are weakly coupled.
Thanks to the presence of the action 
$-\bar{M}^2 \int \rd^4 x \sqrt{-\bar{g}}\bar{R}$, 
the scalar field acquires a kinetic term without the ghost. 
We note that 4DEGB gravity is plagued by the strong 
coupling problem because of the absence of 
such a healthy kinetic term of the radion field.

In Eq.~(\ref{fieldeq}), we derived the covariant gravitational 
field equations by varying the action (\ref{Sga2}) 
with respect to $\bar{g}_{\mu \nu}$. 
The trace of this equation can be expressed in the simple form 
(\ref{RGre}). Varying the action with respect to $\phi$ 
leads to the scalar field equation $2M^2 R-\alpha {\cal G}+\beta W^2=0$ 
and hence the combination with Eq.~(\ref{RGre}) gives 
$2\bar{M}^2 \bar{R}-\alpha \bar{\cal G}+\beta \bar{W}^2=0$. 
Thus, the GB and Weyl terms explicitly affect the spacetime 
geometry in both frames with the metrics 
$g_{\mu \nu}$ and $\bar{g}_{\mu \nu}$.
So long as the Weyl trace anomaly is absent, i.e., $\beta=0$, 
the resulting theory is equivalent to a subclass of Horndeski theories 
given by the action (\ref{action}) with the coupling functions (\ref{G2345}). 
In this paper, we studied whether the GB trace anomaly induces 
a hairy BH solution without instabilities.

In Sec.~\ref{BHsec}, we first showed that the linearly stable 
BH solution in theories without the GB trace anomaly ($\alpha=0$) 
is restricted to a no-hair Schwarzschild solution given by Eq.~(\ref{Sch}). 
In this case there exists the other hairy BH solution (\ref{Beken}) originally 
found by BBMB for a conformally invariant scalar, but this solution is known 
to be unstable (see Appendix \ref{AppA}). 
For $\alpha \neq 0$, we derived the solutions to $\bar{f} (\bar{r})$, 
$\bar{h} (\bar{r})$, and $\phi (\bar{r})$ expanded with respect to 
a small coupling constant $\alpha$ in the frame with 
the metric $\bar{g}_{\mu \nu}$. 
The BH solution expanded up to second order in $\alpha^i$, 
which is consistent with regular boundary conditions on the horizon 
and at spatial infinity, is given by Eqs.~(\ref{fso})-(\ref{phiso}), 
with the BH scalar charge (\ref{qs}). 

The two metric components $\bar{f}(\bar{r})$ and $\bar{h}(\bar{r})$ 
are not proportional to each other for our BH solution, 
so it is different from the exact solution derived 
by imposing the condition 
$\bar{f}(\bar{r})=c_0 \bar{h}(\bar{r})$ 
(see Appendix \ref{AppB}). 
Indeed, the latter is not consistent with all the 
field equations of motion and regular boundary conditions 
on the horizon. As we observe in Fig.~\ref{fig1}, our 
analytic hairy BH solution (\ref{fso})-(\ref{phiso}), which is 
valid for $\hat{\alpha} \ll 1$, exhibits good agreement 
with the numerical results. 
We also derived the metric components $f(r)$ and $h(r)$ 
in the frame with the metric $g_{\mu \nu}$ as 
(\ref{fsod})-(\ref{hsod}) and computed the fifth force 
exerted on a test particle away from the horizon.

In Sec.~\ref{stasec}, we studied the stability of the hairy BH 
solution (\ref{fso})-(\ref{phiso}) against linear perturbations 
on the background metric (\ref{background}).
In the odd-parity sector, the stability conditions are given 
by Eqs.~(\ref{cGdef})-(\ref{cHdef}).  
Provided that $M^2 e^{2\phi}-\bar{M}^2>0$, they can be 
satisfied for the small coupling constant in the range $\hat{\alpha} \ll 1$.
Under this inequality, the no-ghost condition for even-parity 
perturbations is also consistently satisfied without the 
strong coupling problem.  
The radial and angular propagation speeds of three dynamical 
perturbations in the odd- and even-parity sectors are close to 1 
with small corrections induced by the trace anomaly. 
Thus, there are no Laplacian instabilities for our hairy BH solution 
under the conditions $\hat{\alpha} \ll 1$ and $\bar{M} \ll M$. 
These properties are different from those for exact BH solutions 
known in 4DEGB gravity, which are plagued by the strong coupling 
as well as the linear instability problems \cite{Tsujikawa:2022lww,Kase:2023kvq}.

Since the BH solution obtained in this paper possesses the scalar charge 
(\ref{qs}), it may be possible to probe its signature from 
gravitational waves emitted from the binary system 
containing BHs (see e.g., Refs.~\cite{Will:1994fb,Yunes:2009ke,Higashino:2022izi,Quartin:2023tpl}). 
While we have neglected effects of the Weyl trace anomaly 
on the BH solutions and its stabilities, it may also give rise to an 
additional scalar hair for BHs and possibly for NSs. 
Since the Weyl term gives rise to the field equations of motion 
containing derivatives higher than second order, the BH 
stability conditions derived for Horndeski theories cannot be
literally applied to Weyl gravity theories. 
In the EFT scheme, there should be some way of properly 
dealing with such higher-order derivative terms. 
It will be certainly of interest to study whether 
the linearly stable BH solutions exist or not   
in the presence of full trace anomaly terms.
The detailed studies of such issues are left for future works.

\section*{Acknowledgements}

I thank Gregory Gabadadze for giving me insightful and 
positive comments.
The author is supported by the Grant-in-Aid for Scientific Research 
Fund of the JSPS Nos.~19K03854 and 22K03642.

\appendix

\section{Instability of the BBMB solution}
\label{AppA}

For $\alpha=0$, BBMB originally found an exact BH solution 
with a nonvanishing field derivative in the frame given 
with the metric tensor $g_{\mu \nu}$ \cite{Bocharova:1970skc,Bekenstein:1974sf}. 
In the frame with $\bar{g}_{\mu \nu}$, the BBMB solution 
corresponds to Eq.~(\ref{Beken}), which we will derive from now on.
We will also show that it is unstable against linear 
perturbations for some ranges of $\bar{r}$ outside the horizon.

The BBMB solution has the metric components where 
$\bar{f}(\bar{r})$ is proportional to 
$\bar{h}(\bar{r})$, i.e., $\bar{f}(\bar{r})=c_0\bar{h}(\bar{r})$, 
where $c_0$ is a constant. 
In this case, Eq.~(\ref{baeq3}) is integrated to give 
\be
\bar{f}(\bar{r})=c_0+\frac{C_1}{\bar{r}}+\frac{C_2}{\bar{r}^2}\,,
\ee
where $C_1$ and $C_2$ are integration constants. 
Since we are considering the case $\bar{h}'/\bar{h}=\bar{f}'/\bar{f}$, 
the scalar field satisfies the relation $\phi'^2-\phi''=0$ from Eq.~(\ref{baeq2}). 
Then, we obtain the following integrated solution 
\be
\phi'(\bar{r})=-\ln \left( C_3 \bar{r}+C_4 \right)\,,
\ee
where $C_3$ and $C_4$ are constants.
The other equations of motion are satisfied for
\be
C_2=\frac{C_1^2}{4c_0}\,,\qquad 
C_4=\frac{C_1 C_3}{2c_0}\,,\qquad 
C_1= \pm \frac{2M c_0}{C_3 \bar{M}}\,.
\ee
Setting $C_3=M/(\bar{M}m_0)$, the resulting BH solution is 
given by Eq.~(\ref{Beken}). 

For the exact BH solution (\ref{Beken}), the quantities 
(\ref{cGdef})-(\ref{cHdef}) and (\ref{K}) 
reduce, respectively, to 
\be
\bar{{\cal G}}=\bar{{\cal F}}=\bar{{\cal H}}
=\frac{2\bar{M}^2 \bar{r}(2m_0-\bar{r})}
{(\bar{r}-m_0)^2}\,,\qquad 
\bar{{\cal K}}=\frac{6 \bar{M}^2 m_0^2 \bar{r} (2m_0-\bar{r})}
{(\bar{r}^2-3m_0 \bar{r}+3m_0^2)^2}\,, 
\ee
which are all negative for $\bar{r}>2m_0$. 
Hence the exact BH solution is unstable at the 
distance $\bar{r}>2m_0$. 

On using the transformation properties (\ref{fhr}), the metrics and 
scalar field reduce to the forms
\be
f(r)=\frac{c_0 \bar{M}^2}{M^2} \left( 1-\frac{m}{r} \right)^2\,,
\qquad 
h(r)=\left( 1-\frac{m}{r} \right)^2\,,
\qquad
\varphi=\bar{M} e^{-\phi}=\pm \frac{Mm}{r-m}\,,
\ee
where 
\be
r=m\frac{\bar{r}}{\bar{r}-m_0}\,,\qquad 
m=\frac{\bar{M}}{M} m_0\,.
\label{rm}
\ee
One can choose $c_0=M^2/\bar{M}^2$ by using the time 
reparametrization freedom of $f$, in which case 
$f(r)=h(r)=(1-m/r)^2$. 
On using the first relation of Eq.~(\ref{rm}), 
it follows that the BH instability region $\bar{r}>2m_0$
translates to $r<2m$. Indeed, this result agrees with 
the stability analysis performed in the frame with the metric  
$g_{\mu \nu}$ \cite{Kobayashi:2014wsa}.

\section{Exact solution with $\bar{f}$ proportional to $\bar{h}$}
\label{AppB}

For $\alpha \neq 0$, we derive an exact BH solution by imposing 
the condition $\bar{f}(\bar{r})=c_0\bar{h}(\bar{r})$, 
where $c_0$ is a constant.
Then, from Eq.~(\ref{baeq3}), we obtain the simple differential 
equation 
\be
\bar{h}''=-\frac{2[\bar{M}^2 (2\bar{r} \bar{h}'+\bar{h}-1)
+\alpha \bar{h}'^2]}{\bar{M}^2 \bar{r}^2+2\alpha (\bar{h}-1)}\,.
\ee
This is integrated to give
\be
\bar{h}(\bar{r})=1-\frac{\bar{M}^2 \bar{r}^2}{2\alpha} 
\left[ 1-\sqrt{1-\frac{4\alpha (2m \bar{r}-q)}{\bar{M^2}\bar{r}^4}} \right]\,,
\label{hexa}
\ee
where $m$ and $q$ are integration constants. 
Note that we have chosen the asymptotically flat branch where 
$\bar{h}(\bar{r})$ approaches 1 in the limit 
$\bar{r} \to \infty$, such that 
$\bar{h}(\bar{r})=1-2m/\bar{r}+q/\bar{r}^2+{\cal O}(1/\bar{r}^4)$.
This solution is analogous to the one derived in Ref.~\cite{Fernandes:2021dsb} 
in gravitational theories with a conformal scalar field.
Since we are now imposing the condition $\bar{f}(\bar{r})=c_0\bar{h}(\bar{r})$, 
the right hand side of Eq.~(\ref{baeq2}) vanishes. 
This gives a constraint on the scalar field equation of motion. 
For example, there is a possibility for $\phi$ satisfying 
the relation $\phi'^2-\phi''=0$. 
However, we recall that the hairy BH solution (\ref{fso})-(\ref{hso}) derived 
by using regular boundary conditions for both 
the metric and scalar field on the horizon does not satisfy the relation 
$\bar{f}(\bar{r})=c_0\bar{h}(\bar{r})$. 
This means that imposing the condition 
$\bar{f}(\bar{r})=c_0\bar{h}(\bar{r})$ is not consistent 
with all the field equations of motion and
regular boundary conditions at $\bar{r}=\bar{r}_h$, 
so Eq.~(\ref{hexa}) is not actually a desirable solution to the system.

\bibliographystyle{mybibstyle}
\bibliography{bib}

\end{document}